\documentstyle[epsfig,12pt]{article}

\begin{document}
\topmargin 0pt
\oddsidemargin 5mm

\setcounter{page}{1}
\hspace{6cm} Preprint YerPhI-1506(6)-98, January, 1998
\vspace{2cm}
\begin{center}

{\large{Two-dimensional Stationary Wake Fields in Vortexfree Cold Plasma. 
I}}\\

{\large A.Ts. Amatuni, S.G. Arutunian, M.R. Mailian}\\
\vspace{1cm}
{\em Yerevan Physics Institute}\\
{Alikhanian Brothers' St. 2, Yerevan 375036, Republic of Armenia}
\end{center}

\vspace {5mm}

\section{Introduction}

In one dimensional model of wake field generation by a relativistic 
electron bunch in a cold plasma (the beam-generator is modelled by an 
infinite layer moving in the direction perpendicular to its surface) a 
number of interesting results are obtained. In particular, possibility of 
acceleration of electrons (positrons) by wake fields is shown with the 
acceleration rate proportional to $\gamma_0^{1/2}$ \cite{1}-\cite{2} at 
$n\approx \frac{n_0}{2}$; (see also the references cited in \cite{2}) 
possibility of generation of the fields exceeding the wave breaking limit 
by a special combination of bunches (generator-invertor-dumper) is shown 
in \cite{3}, etc.

However the obtained data require some refinement on the subject of 
finite transverse sizes of bunches, since the bunches of existing and 
designing accelerators and storage rings mainly have longitudinal 
geometry, or some times comparable longitudinal and transverse sizes.

In present paper master system of equations for two-dimensional finite 
charged relativistic bunch (of electrons or positrons) moving in cold, 
collisionless, vortex free, stationary plasma with immobile ions are 
formulated (see also \cite{7}-\cite{8}). The system of nonlinear 
equations in 
partial derivatives is then reduced to the form convinient for 
further numerical analysis.

To check the validity of obtained system of equations two limiting cases 
are considered -for infinite wide (transversal) bunch and for infinite 
long (longitudinal) bunch. Both cases admit exact analytical solutions, 
obtained formely in \cite{1} \cite{2} and \cite{4}. Considered limiting 
cases of the general system as one should expect, coincide with the main 
equations obtained in \cite{2}, \cite{4}. The analysis of solutions 
of mentioned limiting cases from viewpoint of development of numerical 
algorithm for solution of the main general system of equations is also 
presented. Ranges of parameters values are determined, where nonstable or 
nonphysical $(n_e < 0)$ solutions take place,
when doing numerical calculations, in particular, at joining the 
solutions in space regions occupied by a limited bunch and that free from 
it, removal of abovementioned limitations on plasma properties, probably 
will 
promote to the clearing up the nature of  these nonstable or nonphysical 
states.

The third obvious check of the validity of the numerical calculations is 
the passage to the linear limit, which is considered in a number of 
papers (see e.g. \cite{5}, \cite{6}, \cite{8} and references therein).

The work is presented in the three parts. Present paper (part I) is 
devoted to formulation of the system of the master equations of the 
problem and analytical considerations of the limiting cases, which have a 
exact solutions \cite{1}-\cite{6} (transversal and longitudinal 
flat bunches, linear approximation). Presented consideration gives some 
hints to the construction of proper algorithm for numerical calculations.

The second part (II) will be devoted to the construction of the algorithm 
for numerical calculations of the system of basic equations of the 
problem for the general case of the bunch of finite transversal and 
longitudinal dimensions.

The third part (III) will present the results of numerical calculations, 
their analysis and comparison with the existing analytical results.

\section{The Master Equations}

The equation of motion of relativistic plasma electrons in electric 
$\vec{E}$ and magnetic $\vec{H}$ fields is written as:
\begin{equation}
\label{1}
\frac{\partial{\vec{p}}}{\partial{t}}+(\vec{v}\frac{\partial}{\partial 
\vec{r}})\vec{p}=-e\vec{E}-
\frac{e}{c}[\vec{v}\times\vec{H}],
\end{equation}
where $\vec{p}$ is momentum, $\vec{v}$ is the velocity of plasma 
electrons in lab system, 
the electron charge is set $-e$. Introducing scalar $\varphi$ and vector 
$\vec{A}$ potentials for electromagnetic fields through formulas
\begin{equation}
\label{2}
\vec{E}=-grad\varphi-\frac{1}{c}\frac{\partial{\vec{A}}}{\partial{t}}, \quad 
\vec{H}=rotA
\end{equation}
one can represent the equation (\ref{1}) in the form:
\begin{equation}
\label{3}
\frac{\partial}{\partial{t}}(\vec{p}-\frac{e}{c}\vec{A})=
egrad\varphi-mc^2grad\gamma-[rot(\vec{p}-\frac{e}{c}\vec{A})\times\vec{v}],
\end{equation}
where $\gamma$ is the Lorentz-factor of plasma electrons.

One can note, that the eq. (\ref{3}) has a partial solution
\begin{equation}
\label{4}
\quad \vec{p}=\frac{e}{c}\vec{A}+grad\chi,
\end{equation}
\begin{equation}
\label{55}
\quad mc^2\gamma=e\varphi-\frac{\partial{\chi}}{\partial{t}},
\end{equation} 
where $\chi$ is an arbitrary calibrating function. Since from (\ref{4}) 
follows
$rot(\vec{p}-\frac{e}{c}\vec{A})=0$, this solution corresponds to the 
vortex free motion of the plasma electrons considered in \cite{7}, \cite{8}, 
\cite{4}. The eqs. (\ref{4},\ref{55}) along with Maxwell's equations for the 
field 
potentials, with charges and currents corresponding to the motion in plasma 
bunch 
charges (which is supposed to be given-- rigid bunch approximation) 
form a complete system of equations of the cold vortex 
free hydrodynamic plasma model, with immobile ions.

Later on we will work with dimensionless potentials
\begin{equation}
\label{5}
\vec{a}=\vec{A}/(mc^2/e),
\end{equation}
\begin{equation}
\label{66}
f=\varphi/(mc^2/e)
\end{equation}
and we will normalize the density of plasma electrons and bunch charges 
on plasma ions density $n_0$. Introduce also plasma frequency $\omega_p$ 
and wavelength $\lambda_p$
\begin{equation}
\label{6}
\omega_p=(4\pi e^2n_0/m)^{1/2}, \quad \lambda_p=c/\omega_p
\end{equation}
we will use dimensionless coordinates (in units of $\lambda_p$) 
and dimensionless time (in units of $\omega_p^{-1}$); 4-momenta of plasma 
electrons represent in the form 
$(\epsilon,\vec{p}c)=mc^2(\gamma,\vec{\beta}\gamma)$, Lorentz factor and 
velocity of bunch charges denote as $\gamma_0,\vec{\beta}_0c$.

Let's choose a calibrating function $\chi=0$. As a result we obtain a full 
system of equations for the considered plasma-electron bunch model:
\begin{equation}
\label{7}
\gamma=f,
\end{equation}
\begin{equation}
\label{8}
\vec{\beta}\gamma=\vec{a},
\end{equation}
\begin{equation}
\label{9}
\Box{f}+\frac{\partial}{\partial{t}}\left(\frac{\partial{f}}{\partial{t}}+
div\vec{a}\right)=-1+n_e+n_b,
\end{equation}
\begin{equation}
\label{10}
\Box{\vec{a}}-grad\left(\frac{\partial{f}}{\partial{t}}+div\vec{a}\right)=
n_e\vec{\beta}+n_b\vec{\beta}_0,
\end{equation}
where $n_e$ is the plasma electrons density, $n_b$ is the bunch charges 
density (positive for electrons and negative for bunches consisting of 
positively charged particles).

As one can see from relations (\ref{7}) and (\ref{8}) components of 
4-potential are connected on mass surface of plasma electrons:
\begin{equation}
\label{11}
f^2=1+a_y^2+a_z^2
\end{equation}

This follows from the selection of calibration function $\chi=0$; i.e. 
(\ref{11}) is a calibrating condition on potentials. Calibration of 
potentials (\ref{11}) we will call "energetic".

From eqs. (\ref{9}) and (\ref{10}) also follows continuity equation for 
plasma electrons:
\begin{equation}
\label{12}
div(n_e\vec{\beta})+\frac{\partial{n_e}}{\partial{t}}=0
\end{equation}
We are interested in steady state wake field solutions of eqs. 
(\ref{7}-\ref{10}), i.e. only on those in which longitudinal dependence of 
all variables is determined by the longitudinal position of point of 
observation relative to the bunch. Assuming that the bunch propagates 
along the axis z, suppose all the physical variables depending on 
combination $z-\beta_0 t$, which we also will denote as $z$. Let us 
consider the case when all the physical variables depend on transverse 
coordinate $y$ only (flat bunch with horizontal dimensions much more greater 
than vertical one). Then  we obtain three scalar eqs. for three 
values $f, a_y, a_z$ and unknown plasma electrons density function
$n_e$, 
obeying the continuity eq. (\ref{12}):
\begin{equation}
\label{13}
\frac{\partial^2{f}}{\partial{y^2}}+\frac{\partial^2{f}}{\partial{z^2}}-
\beta_0\frac{\partial^2{a_y}}{\partial{z^2}}=-1+n_e+n_b,
\end{equation}
\begin{equation}
\label{14}
\beta_0\frac{\partial^2{f}}{\partial{z}\partial{y}}+(1-\beta_0^2)
\frac{\partial^2{a_y}}{\partial{z^2}}-\frac{\partial^2{a_z}}{\partial{z}
\partial{y}}=n_e\frac{a_y}{f},
\end{equation}
\begin{equation}
\label{15}
\beta_0\frac{\partial^2{f}}{\partial{z^2}}-\frac{\partial^2{a_y}}
{\partial{z}\partial{y}}+\frac{\partial^2{a_z}}{\partial{y^2}}-\beta_0^2
\frac{\partial^2{a_z}}{\partial{z^2}}=n_e\frac{a_z}{f}+n_b\beta_0,
\end{equation}
It is convenient to introduce the new unknown functions
\begin{eqnarray}
\label{16}
v=f-\beta_0a_z,\\ \nonumber
u=\beta_0f-a_z
\end{eqnarray}
Then eqs (\ref{13}-\ref{15}) convert to the form:
\begin{equation}
\label{17}
\frac{\partial}{\partial{y}}\left(\frac{\partial{u}}{\partial{y}}+
(1-\beta^2_0)\frac{\partial{a_y}}{\partial{z}}\right)=-\beta_0+n_e
(1-\beta_0^2)\frac{u}{v-\beta u},
\end{equation}
\begin{equation}
\label{18}
\frac{\partial^2{v}}{\partial{y^2}}+(1-\beta_0^2)\frac{\partial^2{v}}
{\partial{z^2}}=-1+n_b(1-\beta_0^2)+n_e(1-\beta_0^2)\frac{v}{v-\beta u},
\end{equation}
\begin{equation}
\label{19}
\frac{\partial}{\partial{z}}\left(\frac{\partial{u}}{\partial{y}}+
(1-\beta_0^2)\frac{\partial{a_y}}{\partial{z}}\right)=n_e(1-\beta_0^2)
\frac{a_y}{v-\beta u}
\end{equation}

Denote the combination involved in both eqs. (\ref{17}) and (\ref{19}) by
\begin{equation}
\label{20}
\mu \equiv 
\frac{\partial{u}}{\partial{y}}+(1-\beta_0^2)\frac{\partial{a_y}}
{\partial{z}},
\end{equation}
one can symmetrize derivatives in (\ref{17}-\ref{19}) introducing "tilded" 
argument and variables:
\begin{equation}
\label{21}
\tilde{z}=z\gamma_0, \quad \tilde{v}=v\gamma_0, \quad 
\tilde{u}=u\gamma_0,\quad \tilde{\mu}=\mu \gamma_0
\end{equation}

Instead of $n_e$ introduce a modified electron density $N$ by the formula:
\begin{equation}
\label{22}
N=\frac{n_e}{\gamma_0(\tilde{v}-\beta_0 \tilde{u})}
\end{equation}

The system of equations for variables $a_y, \tilde{u}, 
\tilde{v}, \tilde{\mu}, N$ describing considered plasma electron bunch 
model, is written as
\begin{equation}
\label{23}
\frac{\partial^2{\tilde{v}}}{\partial{y^2}}+\frac{\partial^2{\tilde{v}}}
{\partial{\tilde{z}^2}}=\gamma_0(-1+n_b(1-\beta_0^2))+N\tilde{v}
\end{equation}
\begin{equation}
\label{24}
\frac{\partial{\tilde{u}}}{\partial{y}}+\frac{\partial{a_y}}{\partial
{\tilde{z}}}=\tilde{\mu}
\end{equation}
\begin{equation}
\label{25}
\frac{\partial{\tilde{\mu}}}{\partial{y}}=-\beta_0\gamma_0+N\tilde{u}
\end{equation}
\begin{equation}
\label{26}
\frac{\partial{\tilde{\mu}}}{\partial{z}}=Na_y
\end{equation}
\begin{equation}
\label{27}
1+a_y^2+\tilde{u}^2=\tilde{v}^2
\end{equation}

The last equation is the condition of "energetic" calibration of 
potentials (\ref{11}).

One can easily see that in this representation the continuity equation 
(\ref{12}) is resulted from (\ref{25}), (\ref{26}) and has the form
\begin{equation}
\label{28}
\frac{\partial}{\partial{y}}(Na_y)-\frac{\partial}{\partial{\tilde{z}}}
(N\tilde{u})=0
\end{equation}

At numeric modelling of the system of equations (\ref{23}-\ref{27}) one 
of eqs (\ref{25}), (\ref{26}) can be replaced by the eq. (\ref{28}).

One can reduce eqs (\ref{23}-\ref{27}) to eqs. for $a_y, \tilde{v}, 
\tilde{\mu}$. To do so, one can express $N$ e.g. from (\ref{25}) and 
define $\tilde{u}$ from (\ref{27}). Eventually we obtain the system of 
equations
\begin{equation}
\label{29}
a_y\frac{\partial{a_y}}{\partial{y}}-\sqrt{\tilde{v}^2-1-a_y^2}\cdot 
\frac{\partial{a_y}}{\partial{\tilde{z}}}=\tilde{v}\frac{\partial{\tilde{v}}}
{\partial{y}}-\sqrt{\tilde{v}^2-1-a_y^2}\cdot \tilde{\mu}
\end{equation}
\begin{equation}
\label{30}
a_y\frac{\partial{\tilde{\mu}}}{\partial{y}}-\sqrt{\tilde{v}^2-1-a_y^2}\cdot
\frac{\partial{\tilde{\mu}}}{\partial{\tilde{z}}}=-\beta_0\gamma_0a_y
\end{equation}
\begin{equation}
\label{31}
\frac{\partial^2{\tilde{v}}}{\partial{y^2}}+\frac{\partial^2{\tilde{v}}}
{\partial{\tilde{z}^2}}=n_b\gamma_0^{-1}-\gamma_0+\frac{\tilde{v}}
{\sqrt{\tilde{v}-1-a_y^2}}\left(\beta_0\gamma_0+\frac{\partial{\tilde{\mu}}}
{\partial{y}}\right)
\end{equation}

Below we will use bunch models symmetrical with respect to $y$ axis. In 
this case one can easily note, that functions $a_y$ and $\tilde{\mu}$ are 
odd functions on $y, \tilde{v}$ is an even one.

Also note, that nondisturbed plasma (no charged particles bunches and 
wake fields) corresponds to the following values of variables: 
\begin{equation}
\label{311}
\tilde{v}=\gamma_0, \quad \tilde{u}=\beta_0\gamma_0, \quad N=1, \quad 
a_y=\tilde{\mu}=0
\end{equation}

It is also interesting to write the equations in the form when the 
calibrating condition (\ref{27})is an integral of motion of the 
differential equations. This can be done by involving the plasma 
electrons density among the unknown functions. Monitoring of this 
parameter is useful also from the physical viewpoint. Let's write these 
eqs. for deviations of functions $\tilde{u}, \tilde{v}$ and N from the 
vacuum values:
\begin{equation}
\label{355}
\tilde{u}=\beta\gamma+\chi,
\end{equation}
\begin{equation}
\label{366}
\tilde{v}=\gamma+\lambda,
\end{equation}
\begin{equation}
\label{377}
N=1+\nu.
\end{equation}

Using eqs. (\ref{23}-\ref{26}), as well as the continuity eq. (\ref{27}) 
for the values $a_y, \chi, \lambda, \nu$ one can obtain the following eqs.:
\begin{eqnarray}
\label{388}
(1+\nu)(\triangle{a_y}-a_y)=(\beta_0\gamma_0+\chi)\frac{\partial^2\nu}
{\partial{y}\partial{\tilde{z}}}+\nu a_y(1+\nu)-a_y\frac{\partial^2{\nu}}
{\partial{y^2}}-\\ \nonumber
-2\frac{\partial{\nu}}{\partial{y}}\frac{\partial{a_y}}{\partial{y}}+
\frac{\partial{\nu}}{\partial{\tilde{z}}}\frac{\partial{\chi}}{\partial{y}}+
\frac{\partial{\nu}}{\partial{y}}\frac{\partial{\chi}}{\partial{\tilde{z}}},
\end{eqnarray}
\begin{eqnarray}
\label{399}
(1+\nu)(\triangle{\chi}-\chi)=\nu(1+\nu)(\beta_0\gamma_0+\chi)-
(\beta_0\gamma_0+\chi)\frac{\partial^2{\nu}}{\partial{\tilde{z}^2}}+a_y
\frac{\partial^2{\nu}}{\partial{y}\partial{\tilde{z}}}- \\ \nonumber
-2\frac{\partial{\nu}}{\partial{\tilde{z}}}\frac{\partial{\chi}}{\partial 
{\tilde{z}}}+\frac{\partial{\nu}}{\partial{\tilde{z}}}\frac{\partial{a_y}}
{\partial{y}}+\frac{\partial{\nu}}{\partial{y}}\frac{\partial{a_y}}
{\partial{\tilde{z}}},
\end{eqnarray}
\begin{equation}
\label{400}
\triangle\lambda-\lambda=\frac{n_b}{\gamma_0}+\nu(\gamma_0+\lambda)
\end{equation}
\begin{eqnarray}
\label{411}
a_y^2\frac{\partial^2\nu}{\partial y^2}-2a_y(\beta_0\gamma_0+\chi)^2
\frac{\partial^2\nu}{\partial y\partial 
\tilde{z}}+(\beta_0\gamma_0+\chi)^2\frac{\partial^2\nu}{\partial\tilde{z}^2}=
\\ \nonumber
=-(1+\nu)(\gamma_0+\lambda)\frac{n_b}{\gamma_0}+(1+\nu)(-\nu+\gamma_0(\lambda-
\beta_0\chi))+(\beta_0\gamma_0+\chi)(\frac{\partial \nu}{\partial y}
\frac{\partial a_y}{\partial \tilde{z}}+ \\ \nonumber
+\frac{\partial \nu}{\partial \tilde{z}}\frac{\partial a_y}{\partial 
y}-2\frac{\partial 
\nu}{\partial\tilde{z}}\frac{\partial\chi}{\partial\tilde{z}})+a_y(\frac
{\partial\nu}{\partial 
y}\frac{\partial\chi}{\partial\tilde{z}}+\frac{\partial\nu}{\partial\tilde{z}}
\frac{\partial\chi}{\partial y}-2\frac{\partial \nu}{\partial 
y}\frac{\partial a_y}{\partial y})- \\ \nonumber
-(1+\nu)\left[\left(\frac{\partial\lambda}{\partial 
y}\right)^2+\left(\frac{\partial \lambda}{\partial
\tilde{z}}\right)^2-\left(\frac{\partial \chi}{\partial
y}\right)^2-\left(\frac{\partial 
\chi}{\partial\tilde{z}}\right)^2-\left(\frac
{\partial a_y}{\partial y}\right)^2-\left(\frac{\partial a_y}{\partial
\tilde
z}\right)^2\right]
\end{eqnarray}
The calibrating condition is written as:
\begin{equation}
\label{422}
2\gamma_0(\lambda-\beta_0\chi)+\lambda^2-a_y^2-\chi^2=0
\end{equation}
In spite of rather unwieldy form of eqs. (\ref{388}-\ref{411}) they have 
a number of advantages for numerical calculations. Firstly, the 
relationship (\ref{422}) follows from eqs. (\ref{388}-\ref{411}) and it can 
used for verification of calculations. Secondly, from eqs. 
(\ref{388}-\ref{411}) one can easily come to the linear case 
corresponding to 
the bunches with low density. In so doing the linearized eqs. are written 
as:
\begin{equation}
\label{433}
\triangle a_y-a_y=\beta_0\gamma_0\frac{\partial^2\nu}{\partial 
y\partial \tilde{z}}
\end{equation}
\begin{equation}
\label{444}
\triangle 
\chi-\chi=\beta_0\gamma_0(\nu-\frac{\partial^2\nu}{\partial\tilde{z}})
\end{equation}
\begin{equation}
\label{455}
\triangle\lambda-\lambda=\frac{n_b}{\gamma_0}+\gamma_0\nu
\end{equation}
\begin{equation}
\label{466}
\beta_0^2\gamma_0^2\frac{\partial^2\nu}{\partial\tilde{z}^2}=-n_b-\nu
\end{equation}
The calibration relationship
\begin{equation}
\label{477}
\lambda-\beta_0\chi=0
\end{equation}
can be easily obtained from eqs. (\ref{444}-\ref{466}).

The linearized eqs. are also a convinient testing area for numerical 
methods in going to the nonlinear eqs.    

\section{Limiting Cases of the Master Equations}

Let us consider two limiting cases of equations (\ref{23}-\ref{25}): 
previously considered case of bunch with infinite transverse sizes and 
case of bunch with finite transverse and infinite longitudinal sizes. 
One can note, that in both cases from the formulation of the problem 
follows, that there is no plasma electrons flow along the axis $y$, i.e 
$\beta_y=0$ and
\begin{equation}
\label{32}
a_y=0.
\end{equation}
Taking into account (\ref{32}) one can reduce (\ref{29}-\ref{31}) to
\begin{equation}
\label{33}
\tilde{v}\frac{\partial{\tilde{v}}}{\partial{y}}-\sqrt{\tilde{v^2}-1}
\tilde{\mu}=0,
\end{equation}
\begin{equation}
\label{34}
\sqrt{\tilde{v^2}-1}\frac{\partial{\tilde{\mu}}}{\partial{\tilde{z}}}=0,
\end{equation}
\begin{equation}
\label{35}
\frac{\partial^2{\tilde{v}}}{\partial{y^2}}+\frac{\partial^2{\tilde{v}}}
{\partial{\tilde{z}}}=n_b\gamma_0^{-1}-\gamma_0+\frac{\tilde{v}}
{\sqrt{\tilde{v^2}-1}}\left(\beta_0\gamma_0+\frac{\partial{\tilde{\mu}}}
{\partial{y}}\right).
\end{equation}

In the case of infinite transverse size 
there are no dependence of all variables on $y$. This case is described 
by a single equation for $\tilde{v}$:
\begin{equation}
\label{36}
\frac{\partial^2{\tilde{v}}}{\partial{\tilde{z}}}=n_b\gamma_0^{-1}-\gamma_0+
\beta_0\gamma_0\frac{\tilde{v}}{\sqrt{\tilde{v}^2-1}}
\end{equation}

Analoguous equation was obtained previously in \cite{2} and was a subject 
on analysis in a series of papers e.g. \cite{7},\cite{8}, \cite{3}. Note, 
that condition (\ref{6}) of absence of vortex 
$$rot(\vec{p}-\frac{e}{c}\vec{A})=0$$ is fulfilled automatically.

If substitute $\tilde{v}$ and $\tilde{z}$ by $v$ and $z$ from (\ref{21}), 
then eq. (\ref{36}) coincides with eq. (\ref{6}) in \cite{3}, having an 
integral 
\begin{equation}
\label{37}
\epsilon=\frac{1}{2}v'^2+\gamma^2[v-\beta(v^2-\gamma^{-2})^{1/2}]-n_bv,
\end{equation}
which allows to consider equation of motion of plasma electrons and 
Maxwell's eqs (Coulomb's Law) as an equation of motion of a point with a 
unit mass, coordinate $v$, velocity $v'$, moving in potential
\begin{equation}
\label{38}
U=\gamma_0^2[v-\beta_0(v^2-\gamma^{-2})^{1/2}]-n_bv.
\end{equation}

\epsfig{file=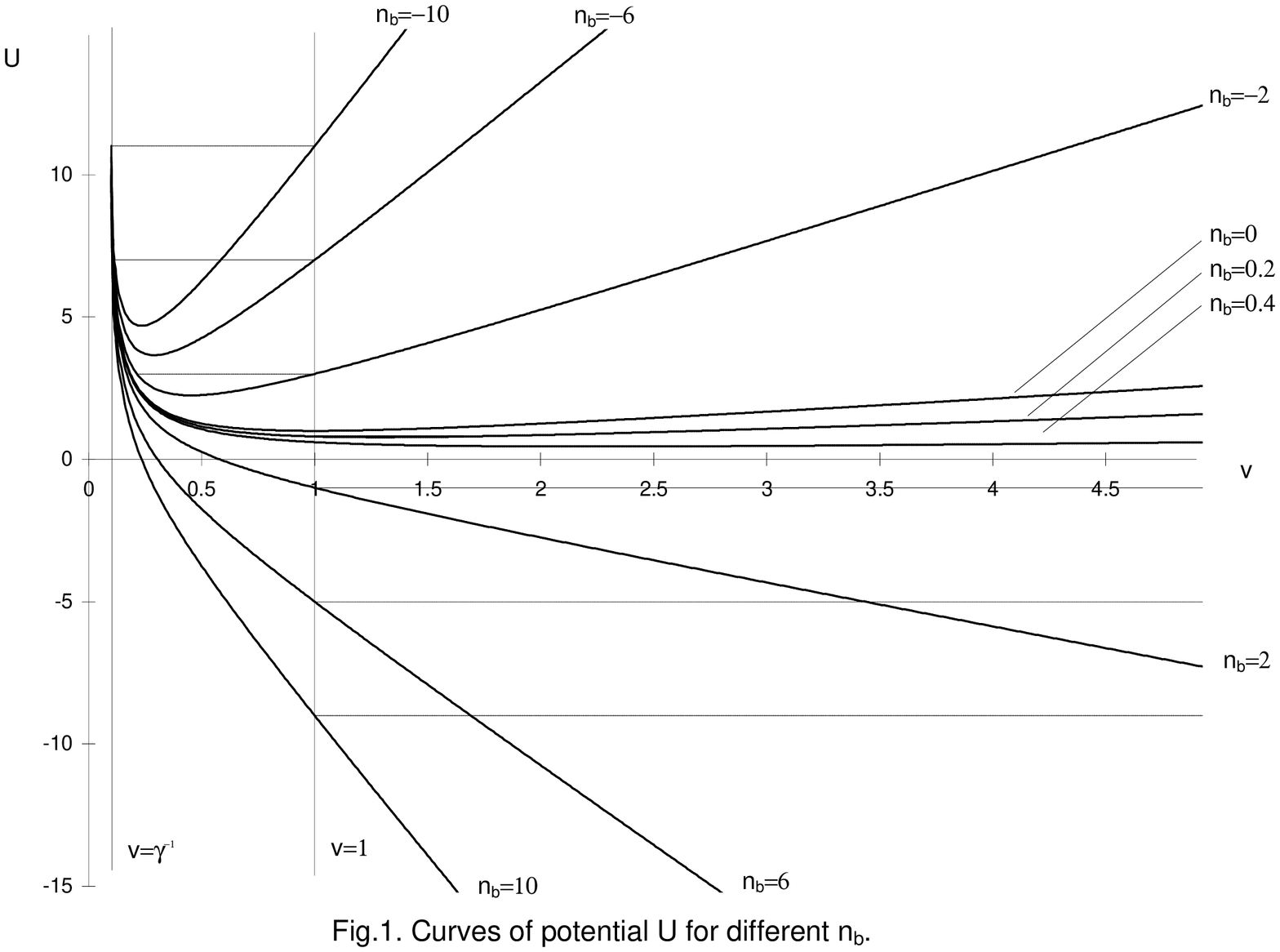,height=15cm,width=10cm,angle=270}

On Fig. 1 $U$ is represented as function of $v$ for different values 
$n_b$ and $\gamma_0=10$. Negative values of $n_b$ correspond to bunches of 
positively charged particles. Boundary values $v=1, U'=0, \epsilon=1-n_b$ 
are reached at front of bunch $z=d$. One can see from Fig. 1, that for 
$n_b \le 0$ the motion is always periodical, it is also periodical for $0 
\le n_b \le 1/2$ and is nonperiodic for $n_b >1/2$. 

\epsfig{file=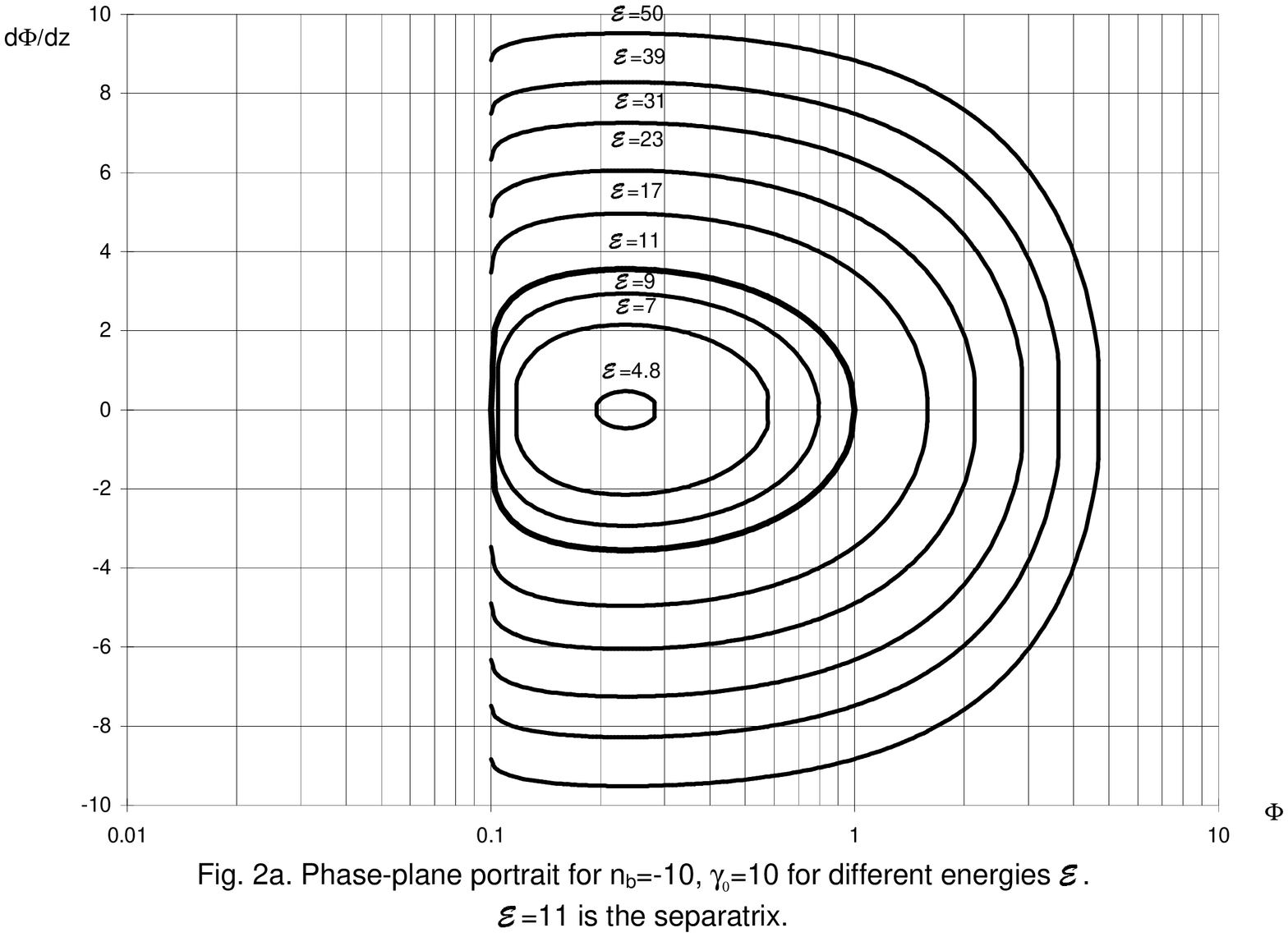,height=9cm,width=7cm,angle=270}
\epsfig{file=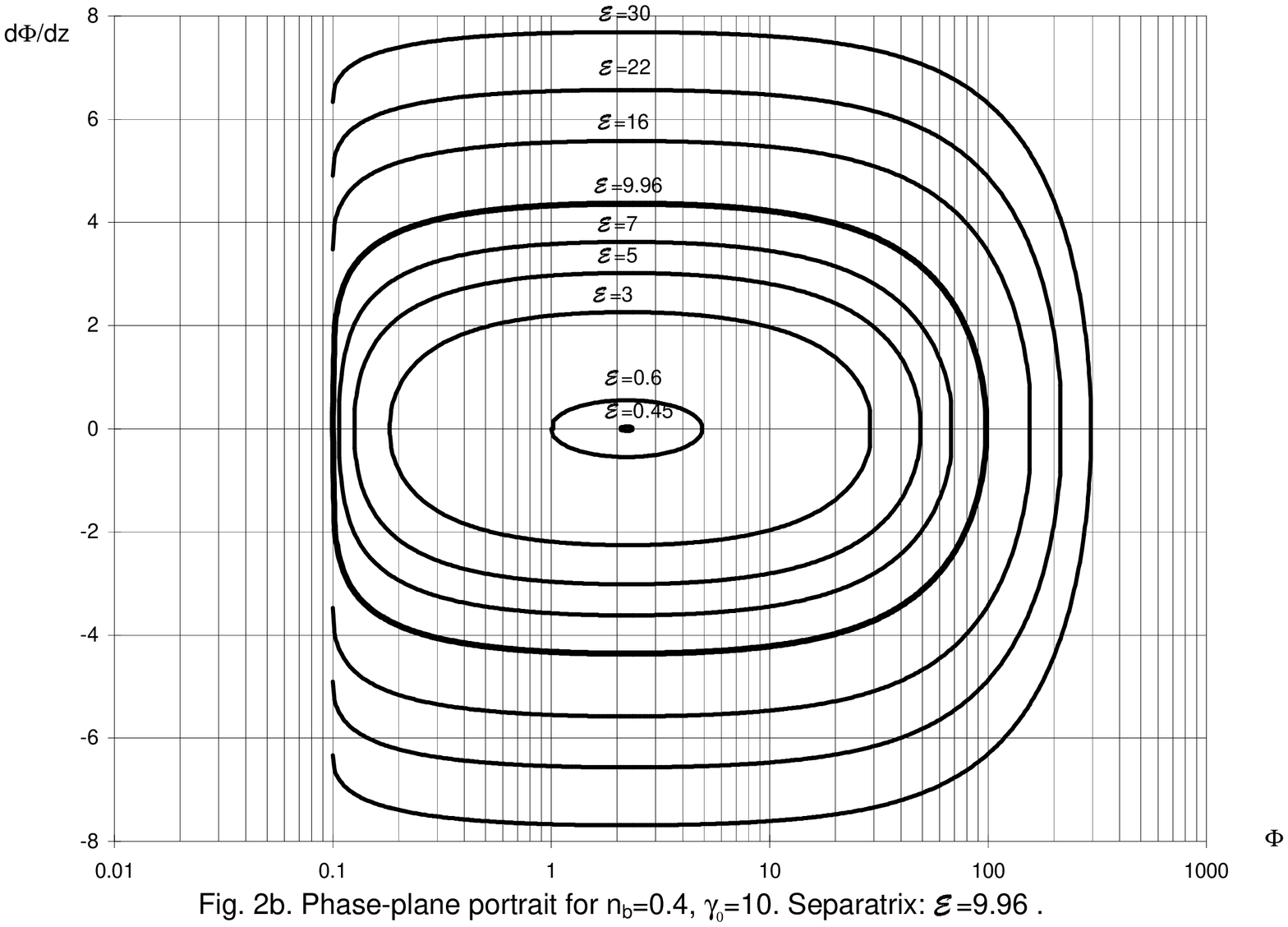,height=9cm,width=7cm,angle=270}
\epsfig{file=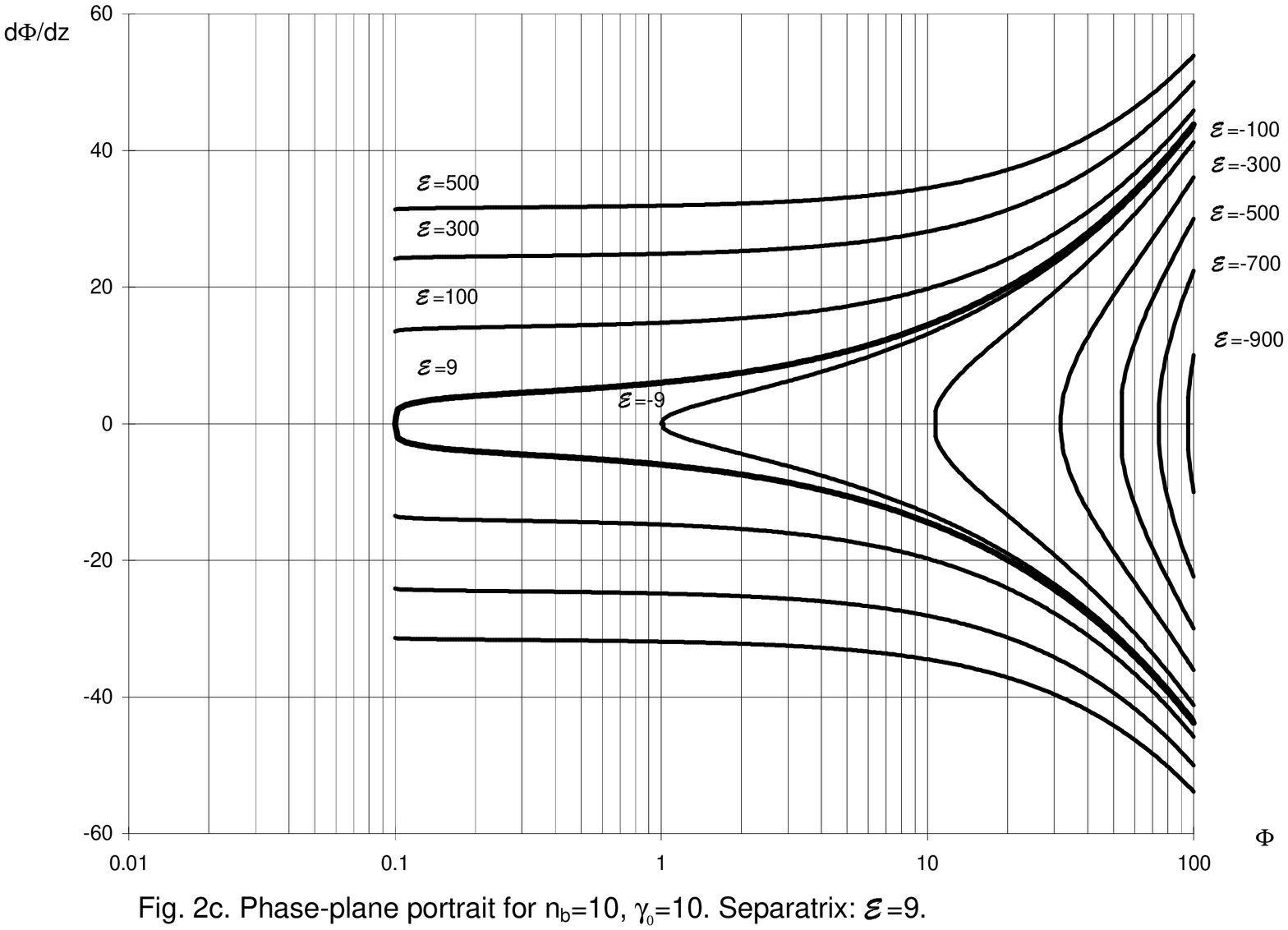,height=10cm,width=8cm,angle=270}

On Fig. 2 phase 
space portraits of system cold plasma -one dimensional flat bunch of 
length $d$ with infinite transverse sizes, $n_b=-10; 0.4; 10, \gamma=10$;
boundary values are not fixed (different $\epsilon$) are presented. One 
can use these trajectories, in particular, to form solutions for combined 
bunches moving in plasma and generating no wake fields \cite{3}. Solution 
for a bunch with given $d, n_b, \beta_0$ must be joined with wave 
solution for a free plasma beyond the bunch (wake waves with $n_b=0$, 
phase velocity of waves coincides with velocity of the bunch $v_{ph}=v_0$
\cite{1}, \cite{2}). In doing so no additional limitations arise if 
$n_b<0$, or $0 \le n_b \le\frac{1}{1+\beta_0}$. For nonperiodical 
solutions inside bunch, when $n_b >\frac{1}{1+\beta_0}, n_e$ may become 
negative; it remains positive for values of dimensionless pulses 
$\rho(0)=p_{ez/mc}$ of plasma electrons on the rear edge of the bunch
$-\beta_0\gamma_0\le\rho(0)\le 0$, i.e. for certain bunch lengths 
$0 \le d \le d_{cr}, \rho(0)$ is limited from above. Negative 
$n_e$ usually interpreted a consequence of violation of steady state 
condition(see e.g. \cite{2}, \cite{3});  in particular, it is possible 
that the wake field will break behind the bunch if the wave amplitude is 
large enough imidiately behind the bunch. 

In the other (longitudinal) limiting case there is no dependence of all 
variables on $\tilde{z}$. Physically such conditions can be realized for 
long enough bunches far from the bunche's head. The linearized problem 
for semi-infinite bunch in cylindrical geometry was considered in 
\cite{12}. A tensor of dielectric permeability was introduced to connect 
the plasme electrons current density with electric field. The essential 
contribution was defined by concomitant fields. The transition fields 
arising at injection decrease exponentially. In the case when the bunch 
transverse size is much smaller than the plasma wave length, a return 
current, concentrated in the bunch transverse section, arizes. In the 
other limiting case, the transverse section of the return current is much 
greater than the bunch transverse section.

From (\ref{33}-\ref{35}) we 
obtain \begin{equation}
\label{39}
\frac{\partial^2{\tilde{v}}}{\partial{y^2}}=n_b\gamma_0^{-1}-\gamma_0+
\frac{\tilde{v}}{\sqrt{\tilde{v^2}-1}}(\beta_0\gamma_0+\frac{\partial^2}
{\partial{y^2}}\sqrt{\tilde{v^2}-1})
\end{equation}

It is interesting to note, that eq. (\ref{39}) (the longitudinal limiting 
case) differs from (\ref{36}) (the transverse  limiting case) by the term
$\partial^2\sqrt{\tilde{v}^2-1}/\partial{y}^2$ in brackets in the right 
side of eq. (\ref{39}).

By substitution
\begin{equation}
\label{40}
\sqrt{\tilde{v}^2-1}=shR, \tilde{v}=chR
\end{equation}
we obtain the following equation for R
\begin{equation}
\label{41}
R''=-(n_b\gamma_0^{-1}-\gamma_0)shR-\beta_0\gamma_0chR
\end{equation}
Prime in (\ref{422}) means derivative with respect to $y$. Note that by 
substitution \begin{equation}
\label{42}
R=ln\gamma_0(1+\beta_0)-\chi
\end{equation}
one can obtain following equation for $\chi$:
\begin{equation}
\label{43}
\chi''=n_b\beta_0 ch\chi+(1-n_b)sh\chi
\end{equation}
This equation is obtained in work \cite{4}, dedicated to the 
transportation of longitudinally infinite bunches of charged particles 
with current, exceeding the Alfven limit through a cold vortexless plasma.

The first integral of (\ref{43}) when $n_b=const$ is 
\begin{equation}
\label{44}
E=\frac{1}{2}(R')^2+(n_b\gamma^{-1}-\gamma_0)chR+\beta_0\gamma_0shR=const
\end{equation}

Thus in this case also the problem is reduced to the study of massive 
(unit mass) point motion in potential field $U(R)$:
\begin{equation}
\label{45}
U(R)=\gamma[\beta_0shR-(n_b\gamma_0^{-2}-1)chR]
\end{equation}
Form of function $U(R)$ essentially depends on the regions of parameter 
values $n_b$ and $\beta_0$.

When
\begin{equation}
\label{46}
n_b > \frac{1}{1-\beta_0},
\end{equation}
the function $U(R)$ has a single minimum and  form the potential 
well.

If
\begin{equation}
\label{47}
n_b < \frac{1}{1+\beta_0},
\end{equation}
the function $U$ has one maximum and looks like a "hill".

In region of values
\begin{equation}
\label{48}
\frac{1}{1+\beta_0} < n_b < \frac{1}{1-\beta_0},
\end{equation}
the function has no extrema and has a characteristic bend, and varies 
from $-\infty$ to $+\infty$.

\epsfig{file=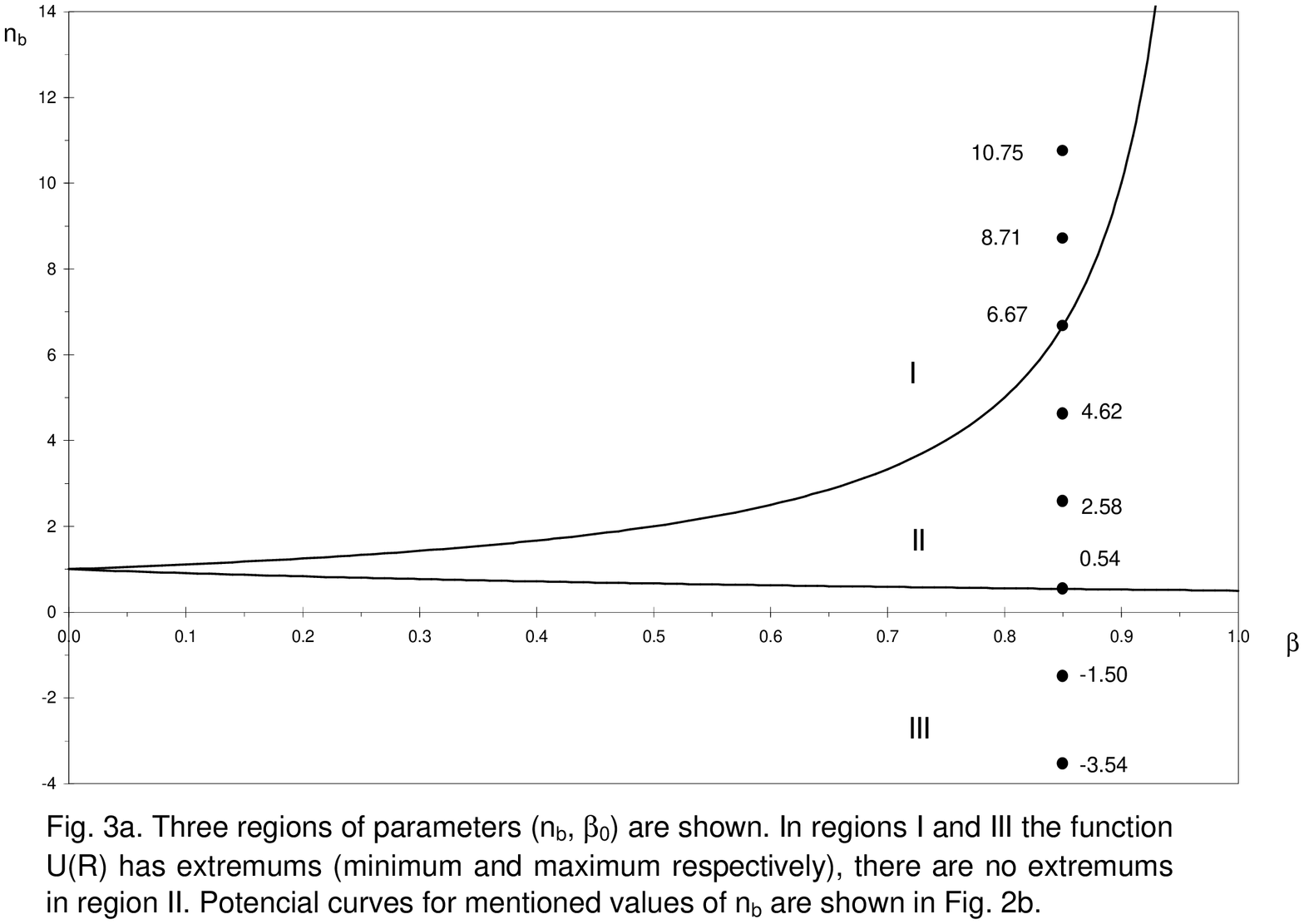,height=9cm,width=8cm,angle=270}
\epsfig{file=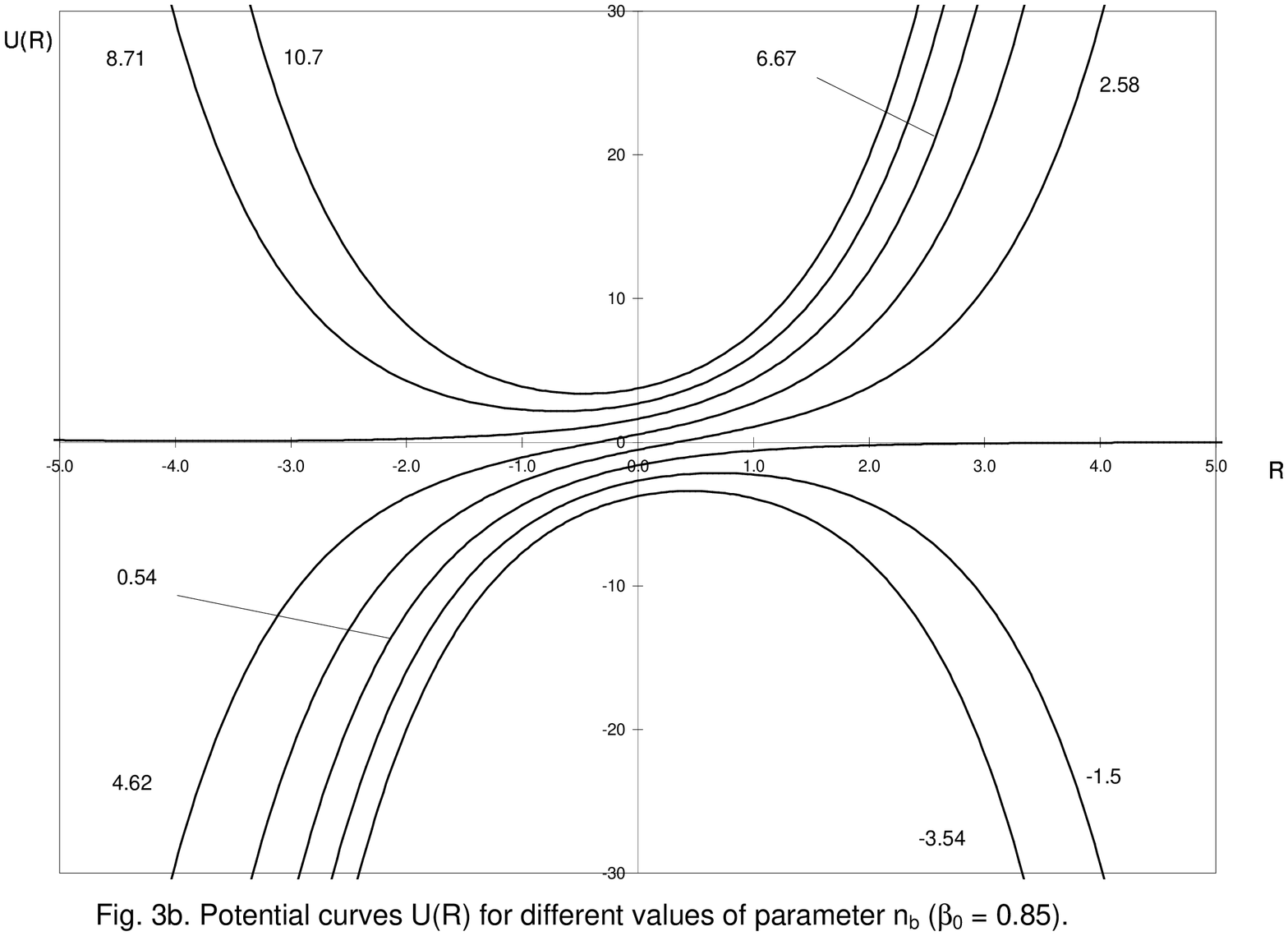,height=9cm,width=8cm,angle=270}

In the fig. 3 region of parameters $(n_b,\beta_0)$ and some potential 
curves for $\beta_0=0.85$ and a few values of $n_b$ are represented.

Phase trajectories in the plane $(R,R')$, corresponding to parameters 
$(n_b,\beta_0)$ from regions I $(n_b > 1/(1-\beta_0))$, II 
$(1/(1-\beta_0) >n_b > 1/(1+\beta_0))$, III $(1/(1+\beta_0) > n_b)$ are 
presented in figs. 4-6.

\epsfig{file=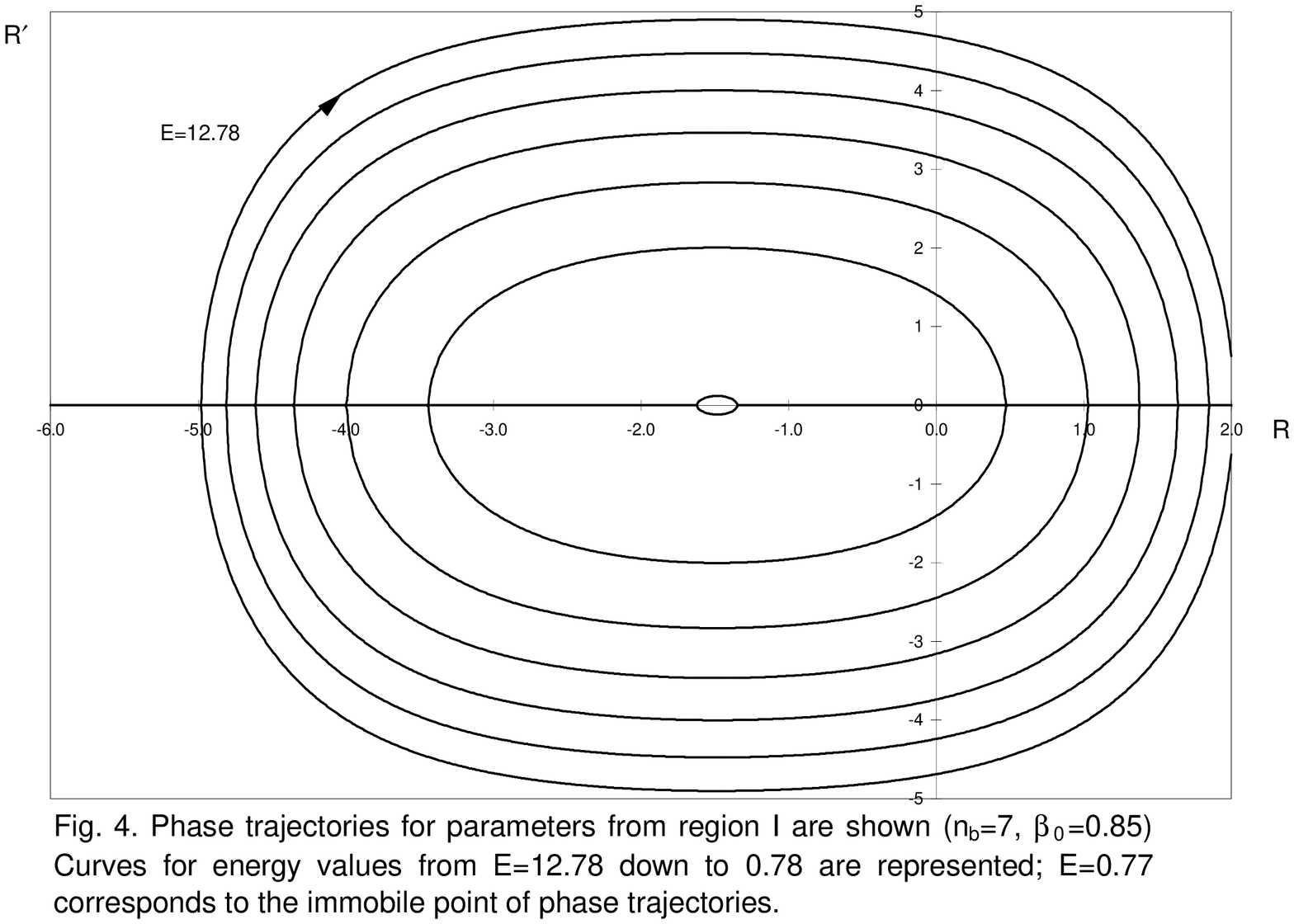,height=9cm,width=9cm,angle=270}
\epsfig{file=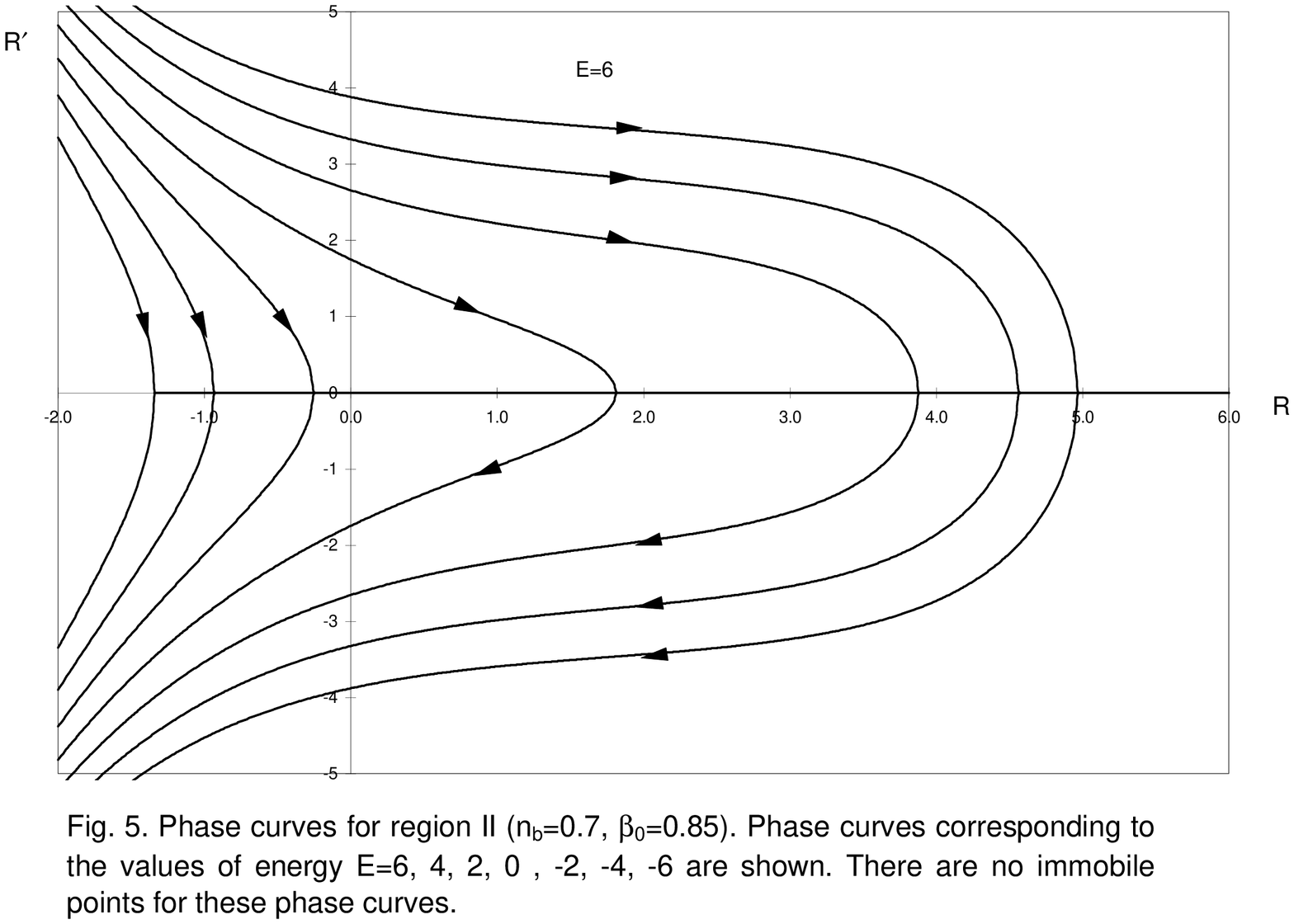,height=9cm,width=9cm,angle=270}
\epsfig{file=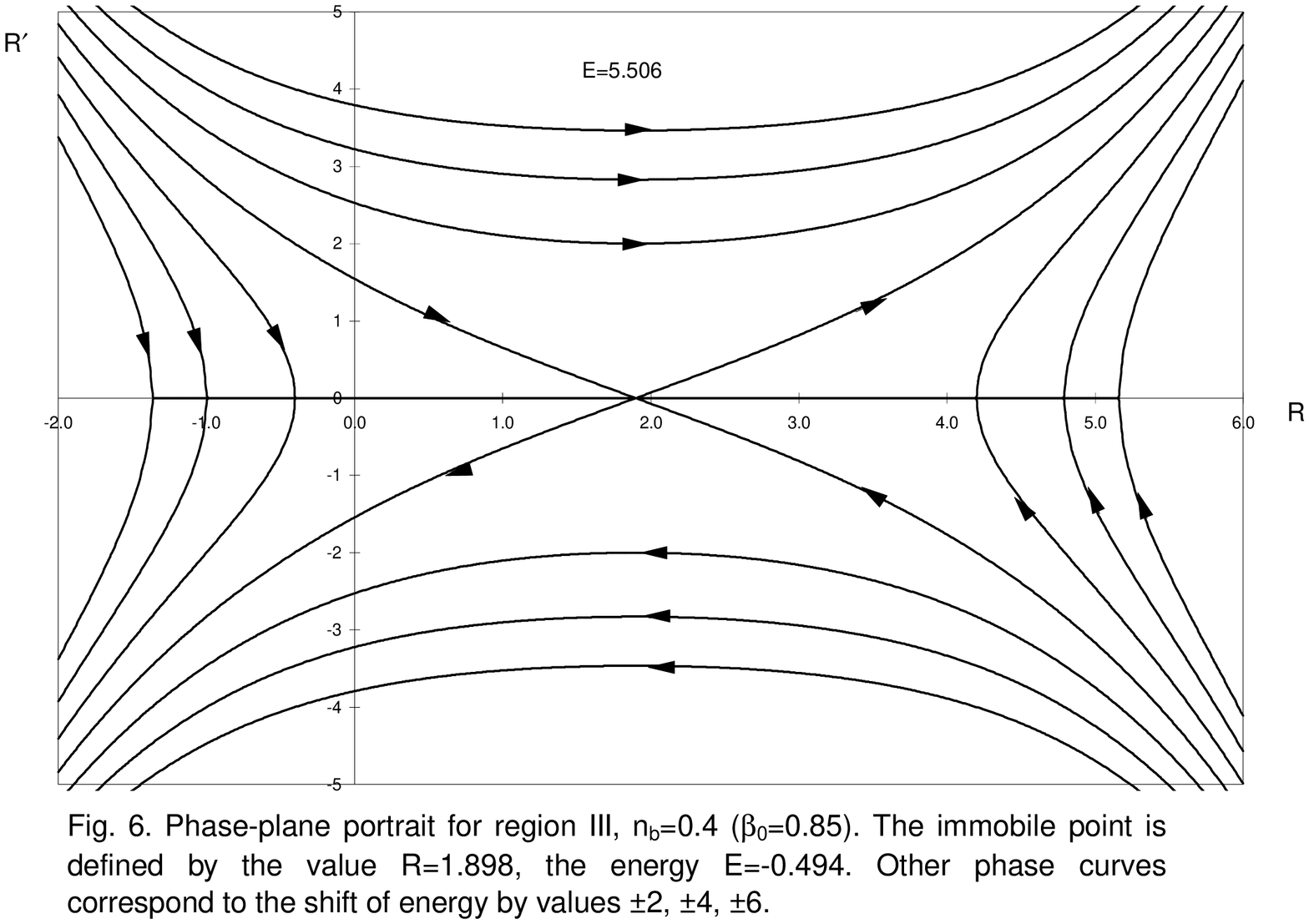,height=9cm,width=9cm,angle=270}

For the region I the phase trajectories are closed and there is a single 
stable immobile point:
\begin{equation}
\label{49}
R'_0=0, \quad R_0=arcth\frac{\beta_0}{1-n_b\gamma_0^{-2}}
\end{equation}

In the region II all phase trajectories are not closed, there are no 
singular points, in the region III only hyperbolic motion around the 
point (\ref{49}) can be realized; the point (\ref{49}) in this case 
becomes unstable one.

Location of $R_0$ for different values of parameters $n_b$ are shown in 
fig. 7.

\epsfig{file=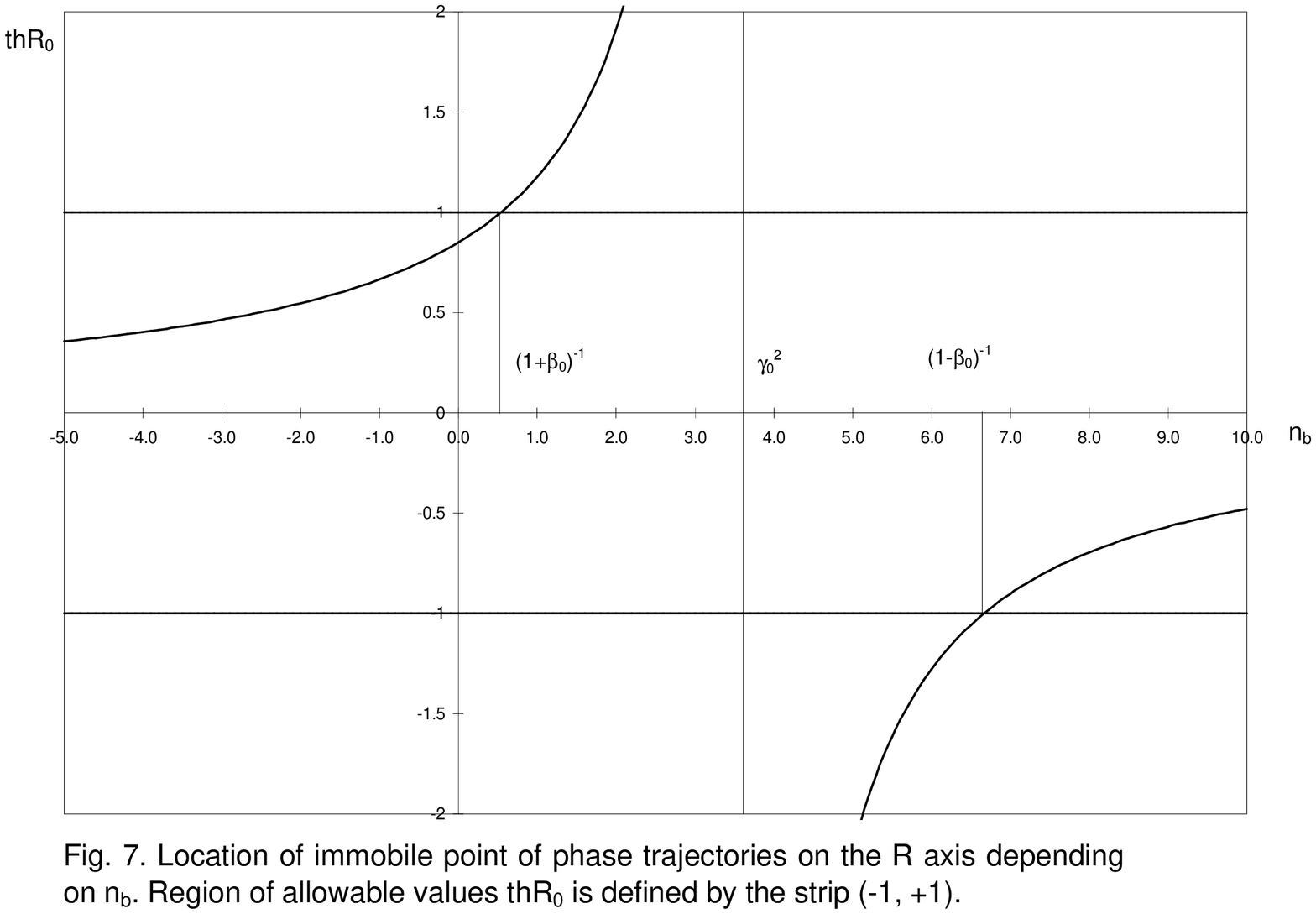,height=12cm,width=12cm,angle=270}

Since finally we will be interested in restricted in transverse direction 
bunches ($n_b=0$ at $|y|>b, 2b$ is the width of flat bunch), the case 
$n_b$=0 requires an additional consideration. Passing from the region 
occupied by bunch $(|y| \le b)$ to the region without it needs the physical 
solutions, corresponding to the null value of all physical quantities 
(electromagnetic fields, pulses of plasma electrons) at $y\rightarrow \pm 
\infty$. These are the solutions which must be joined with the solutions 
inside the bunch, where $n_b \ne 0$. For hyperbolic phase trajectories 
(corresponding to the case $n_b=0$) the separatrices passing through the 
unstable (hyperbolic) immobile point are the only suitable trajectories. 
Passing along two branches of separatrices requires infinite long 
"time-development" on parameter $y$.

Eqs. for the branches of separatrices are:
\begin{equation}
\label{50}
R'=\pm \left[\left(\frac{1-\beta_0}{1+\beta_0}\right)^{1/4}e^{R/2}-
\left(\frac{1+\beta_0}{1-\beta_0}\right)^{1/4}e^{-R/2}\right]
\end{equation}

Construction of solutions, corresponding to the bunches finite with 
respect to $y$, can be done in the following way. Branches of separatrix 
$n_b=0$, realizing the motion in $y$ from the side surfaces of the bunch 
to the infinity are to be superimposed on phase trajectories 
corresponding to the position of $(\beta_0,n_b)$ in one of the regions 
I,II, III. Separatrix branches intersect many phase 
trajectories-different choices of intersection regions correspond to the 
different widths of bunches with the same values of parameters 
$(n_b,\beta_0)$. Figs. 8 and 9 represent the process for construction of 
solutions in regions II and III.

\epsfig{file=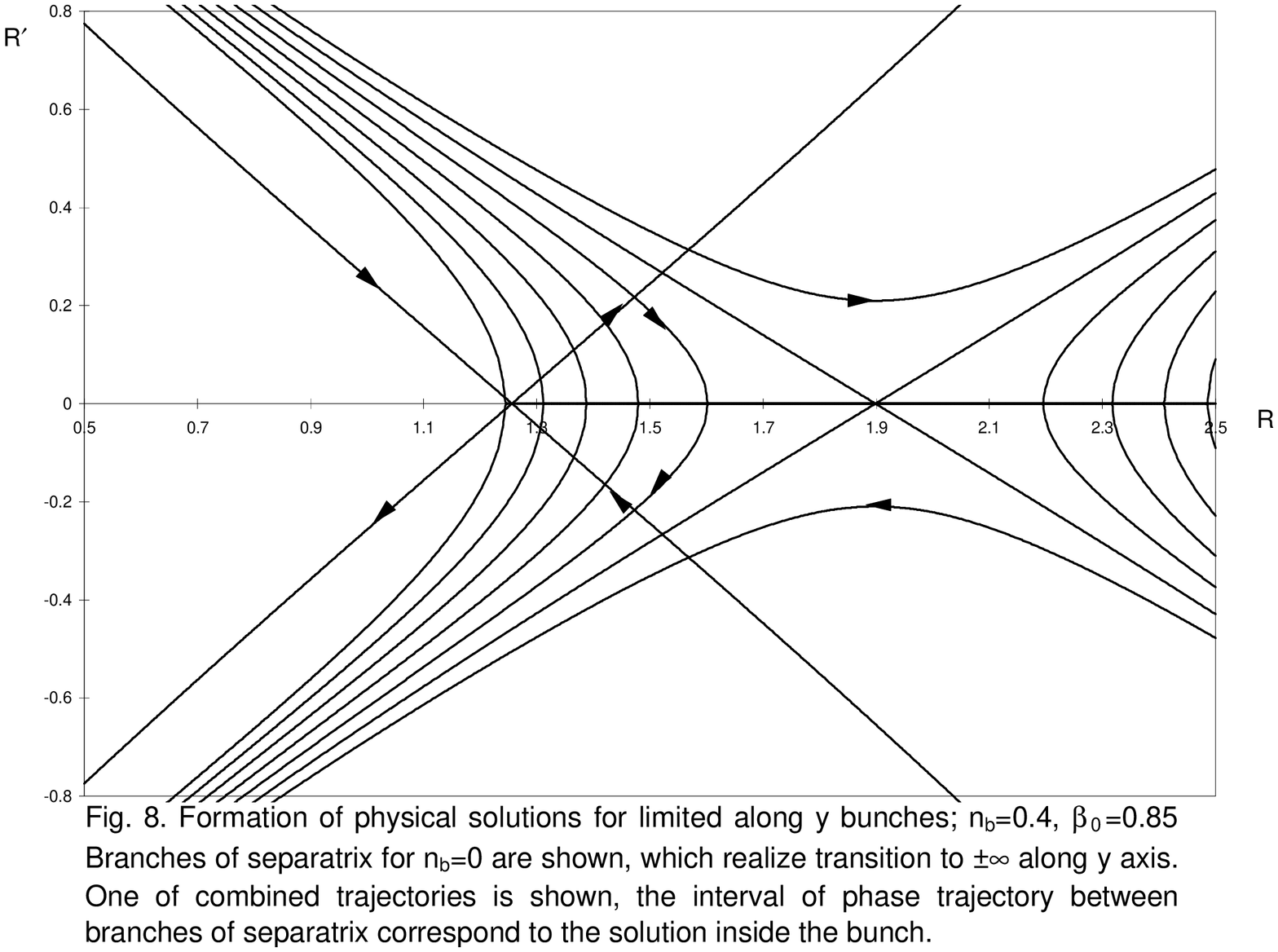,height=9cm,width=9cm,angle=270}
\epsfig{file=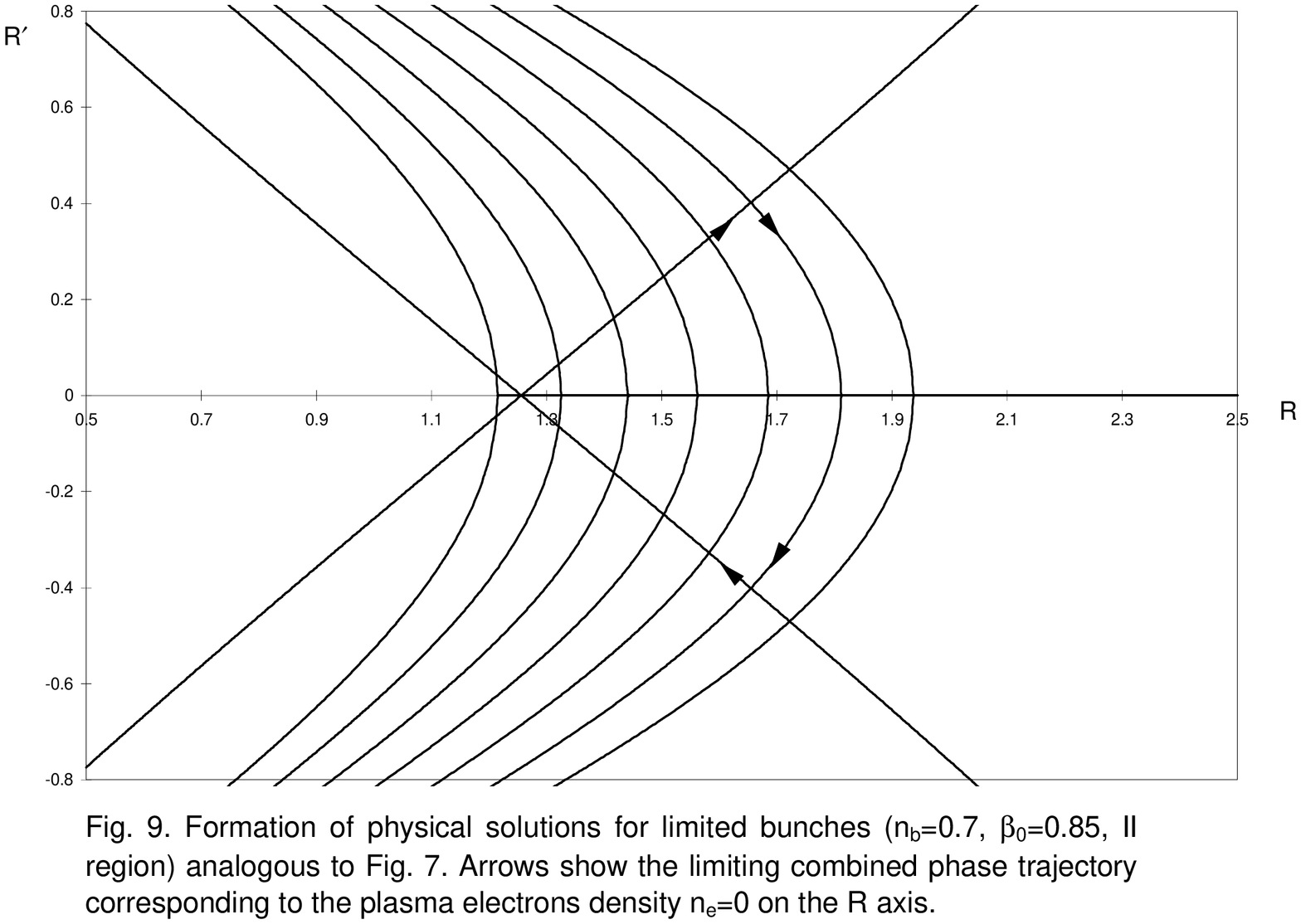,height=9cm,width=9cm,angle=270}

Note an important circumstance. There are no transverse motion of plasma 
electrons in adopted model and adjacent layers of plasma slide one with 
respect to 
other (because we neglect the plasma viscosity). In such a formulation 
the continuity equation is satisfied automatically and from the 
mathematical viewpoint there are solutions with any values of $n_e$, 
including negative ones, found, in particular in \cite{2}, \cite{4}.

However, clearly such solutions are not physical and are to be discarded.
In \cite{4} an illegitimate, (by our opinion), attempt is done to interpret 
the regions with negative $n_e$ as channels from which all plasma 
electrons are displaced.

Using the formula (\ref{22}) and relation (\ref{44}) one can obtain the 
following expression for plasma electrons density:
\begin{equation}
\label{51}
n_e=\gamma_0chR(1-\beta_0thR)(-E+\frac{3}{2}(R')^2).
\end{equation}

From (\ref{51}) follows that $n_e$ can alter its sign, and it can take 
place only if $E > 0$. Indeed, as it is seen from construction of 
physical solutions by means of joining of phase trajectories for $n_b \ne 
0$ with branches of separatrix at $n_b = 0$, such a combined phase 
trajectory always intersects the $R$ axis $(R'=0)$. Thus, for positive 
$E$ inside the bunch $n_e <0$. It is clear, that there are no 
physical solutions when $n_b>1/(1-\beta_0)$ (region I of values 
$(n_b,\beta_0)$), since the minimum values of corresponding potential 
pits are positive. In contrary, in the region III $(n_b <1/(1+\beta_0))$ 
the condition $E<0$ does not impose any additional limitations, since the 
regions for joining of solutions correspond to $E<E_0$, where $E_0<0$ is 
the maximum value of potential $U(R)$.

Lastly, in the II region $(1/(1-\beta_0)>n_b>1/(1+\beta_0))$ intervals 
of phase trajectories necessary for joining with branches of the immobile 
point of phase trajectories correspond to the energy values $E>0$, when 
$n_b>1$. Therefore at $1/(1-\beta_0)>n_b>1$ inside bunch always $n_e <0$. 
Such solutions as it was mentioned above are not physical by our 
opinion. In the region $1>n_b>1/(1+\beta_0)$ the condition $n_e>0$ is not 
satisfied for a limited set of values of energy only $(n_b-1 <E <0)$. The 
maximum allowed bunch width corresponds to the "time" which is necessary 
for passing the section of phase trajectory at $E=0$ between the branches 
of separatrix, where $n_b=0$.

In fig. 9 arrows indicate the limiting combined phase trajectory 
corresponding to the plasma electrons density $n_e=0$ in the middle of 
bunch. Phase trajectories right to the limiting one correspond to the 
nonphysical $(n_e < 0)$ solutions.

Complete sweeping of plasma electrons by the bunch, i.e. formation of a 
channel in plasma take place at $(R')^2=\frac{2}{3}E \ge 0$. In 
particular, this condition, as it was mentioned, is satisfied in the 
middle of bunch $(R'=0)$ at $E=0$, i.e. at 
$R=arcth(\beta_0/(1-n_b\gamma_0^{-2}))$ in an immobile point of the region 
I, (which is unstable for region III).

Thus, summarising the contents of the presented consideration one can 
note, that for a transversely infinite bunch at $n_b < 1/(1+\beta_0)$ 
physical solutions exist for bunches of any width. In the case 
$1/(1+\beta_0)<n_b<1$ physical solutions exist for bunches of limited 
width only. When $n_b>1$ in the middle of bunch $n_e<0$ and we discard 
such a solutions as nonphysical ones for considered formulation of the 
problem (collisionless, stationary, vortexless, cold plasma).

Very likely that consideration of the problems which takes into account 
violation of stationary condition, thermal motion of plasma 
electrons and ions, plasma viscosity possibly will allow to clear out the 
physical nature of these nonstable or "nonphysical" solutions.

It is necessary to remember that states with $n_e <0$ are obtained in 
very long or very wide bunches. The finite dimensions of real bunches can 
change the domain of existance of these states or even eliminate them 
completely. Linear approximation to the limiting cases of formulated 
problem, discussed particularly in \cite{5}, \cite{6}, \cite{8}, is 
valid, when bunch density is $n_b \ll 1$, and plasma electrons are 
nonrelativistic $|\vec{\beta}| \ll 1$. The last assumption provide that 
the condition $rot (\vec{p}-\frac{e}{c}\vec{A})=0$ for the absence of the 
vortexes in plasma flow is fulfilled automaticaly, due  to nonrelativistic 
equation of motion of plasma electrons and Faraday's law \cite{8}. 
Condition $n_b \ll 1$ allows to seek the plasma electron density as 
$n_e=1+n_e', |n_e'| \ll 1$ and linearize the continuity eq. (\ref{12}) 
and then, using nonrelativistic equation of plasma electron motion and 
Coulomb law, find a solution for $n_e'$. According to the condition 
$\beta \ll 1, f$ in (\ref{7}) in linear approximation is $f=1+f', |f'| 
\ll 1, \vec{a} \approx \vec{\beta}$ and eqs. (\ref{9}), (\ref{10}) with 
the given right side have exact analytical solution, which can served as 
a test function for subsequent limiting case in computer simulation of 
the general problem. It is necessary to take into account that condition 
$\beta \ll 1$, can be fulfiled only for short enough bunches.

Obtained results of presented analytical consideration of general system 
of equations for two-dimensional limited bunch of charged particles 
moving in a cold, collisionless, vortexless, stationary cold plasma, as 
well as the limiting cases corresponding to the infinitely wide 
(transverse) and infinitely long (longitudinal) bunches and linear 
approximation, valid for $n_b \ll 1, \beta \ll 1$, must serve as an 
analytical bases for formulation of an algorithm for numerical solution 
of the problem-obtaining of wake fields and focusing forces for bunches 
of an arbitrary longitudinal and transverse sizes.

\end{document}